\documentclass[fleqn,twoside]{article}
\usepackage{espcrc2}

\usepackage{graphicx}
\usepackage[figuresright]{rotating}

\newcommand{\beq}{\begin{equation}}
\newcommand{\eeq}{\end{equation}}
\newcommand{\tr}{\mbox{Tr}}
\newcommand{\eq}[1]{(\ref{#1})}

\newcommand{\AmS}{{\protect\the\textfont2
  A\kern-.1667em\lower.5ex\hbox{M}\kern-.125emS}}

\title{
\vspace{-9mm} \rightline{\small ITEP-LAT/2003-23} \vspace{-2mm}
\rightline{\small September, 2003}
Interplay of monopoles and P-Vortices
    \thanks{Talk presented by S.N.S. 
	at Lattice 2003, Tsukuba
    }
}

\author{A.V.~Kovalenko\address[ITEP]{Institute of Theoretical and
Experimental Physics, B.~Cheremushkinskaya~25, Moscow, 117259, Russia},
M.I.~Polikarpov\addressmark, S.N.~Syritsyn\addressmark~and
V.I.~Zakharov\address[MPI]{Max-Planck Institut f\"ur Physik, F\"ohringer Ring
6, 80805, M\"unchen, Germany}\thanks{Work is partially supported by grants RFBR
02-02-17308, RFBR 01-02-17456, DFG-RFBR 436 RUS 113/739/0, INTAS-00-00111 and
CRDF award RPI-2364-MO-02.} }
\begin{document}

\begin{abstract}
We show that P-Vortices in the confinement phase of SU(2) lattice gauge theory
form one large percolating (infrared) cluster and a number of small
(ultraviolet) clusters. We discuss the interrelation of clusters of monopoles
in the maximal Abelian projection with clusters of P-vortices. To extract
P-vortices we use both direct and indirect central projections and find  
qualitatively similar results.
\end{abstract}

\maketitle

\section{INTRODUCTION}
Both the monopole and P-vortex mechanisms of the confinement
are supported  by results of lattice simulations (for review see, e.g.,
Refs.~\cite{reviews}). Thus, both the monopoles and P-vortices
appear to be adequate dynamical variables to describe the infrared
physics. Then one could expect that they are in fact
interrelated. Here we will address this issue in terms of the geometry of
the monopole and vortex clusters. The results, in principle, could
depend on the projection used. To define the P-vortices we will use
both direct  maximal center projection (DMCP)~\cite{greens2} and
indirect maximal center projection (IMCP)~\cite{IMC}. DMCP in SU(2) lattice
gauge theory is defined by maximization, with respect to 
gauge transformations of the functional
\beq
F(U) = \sum_{n,\mu} \left( \tr U_{n,\mu} \right)^2 \, , \label{maxfunc}
\eeq
where $U_{n,\mu}$ is the lattice gauge
field. Condition (\ref{maxfunc}) fixes the gauge up to Z(2) gauge
transformations. The Z(2) gauge fields are defined as: $Z_{n,\mu} = \mbox{sign}
\tr U_{n,\mu}$. The plaquettes $Z_{n,\mu\nu}$ constructed from links
$Z_{n,\mu}$ have values $\pm 1$. The P-vortices (which form closed surfaces in
4D space) are made from the plaquettes, dual to plaquettes with $Z_{n,\mu\nu} =
-1$. To get IMCP we first fix the maximally Abelian gauge 
and extract then monopole currents 
from the Abelian variables $e^{i\theta_{n,\mu}}$. 
Finally, we
project the residual U(1) gauge degrees of freedom into Z(2) by the procedure
completely analogous to the DMCP case, by substitution into
eq.~\eq{maxfunc} of the Abelian
gauge field: $U_{n,\mu} \to U_{n,\mu}^{Ab}\equiv
\mbox{diag}\left( {e^{i\theta_{n,\mu}}, e^{-i\theta_{n,\mu}}}\right)$.

We work at various lattice spacings and extrapolate our results to the
continuum limit. The gauge fields are generated on $16^4$ lattice at
$\beta=2.35$ (20 statistically independent configurations), $24^4$ lattice at
$\beta=2.40$ (50 configurations), $\beta=2.45$
(20 configurations), $\beta=2.50$ (50
configurations), and on $28^4$ lattice at $\beta=2.55$ (40 configurations),
$\beta=2.60$ (50 configurations). To fix the physical scale
we use the string tension in lattice
units~\cite{Fingberg:1992ju}, $\sqrt\sigma = 440\, MeV$. We fix IMCP and
DMCP using the simulated annealing algorithm~\cite{SA}, which is the most
precise method to fix gauge on the lattice.

\section{P-VORTEX CLUSTERS}
It is well known~\cite{HT}, \cite{Boyko} that monopole currents form two types
of clusters: one large (infrared, IR) percolating cluster and many finite
(ultraviolet, UV) clusters. The density of the monopole cluster is defined as:
$\rho = <l>/(N_l a^3)$, where $<l>$ is the average number of links in the
monopole cluster, $N_l = 4L^4$ is the total number of links on the lattice. The
density of the IR monopole cluster scales as a physical quantity ($\rho_{\rm
IR}^{\rm mon}~=~7.70(8)\, fm^{-3}$), while the density of ultraviolet monopoles
diverges near the continuum limit as $1/a$, where $a$ is the lattice
spacing \cite{Boyko}, \cite{BornMP}.
\begin{figure}[h]
\vspace{-5mm}
\includegraphics[scale=0.30,angle=270]{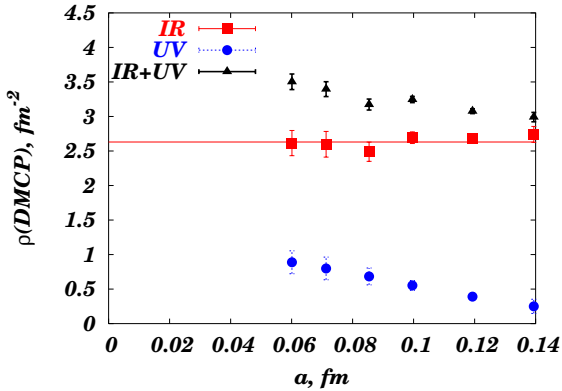}
\vspace{-5mm}
\caption{P-Vortex densities in DMCP.
    \label{fig:PVD_scaling}}
\vspace{-5mm}
\end{figure}
\begin{figure}[h]
\vspace{-5mm}
\includegraphics[scale=0.30,angle=270]{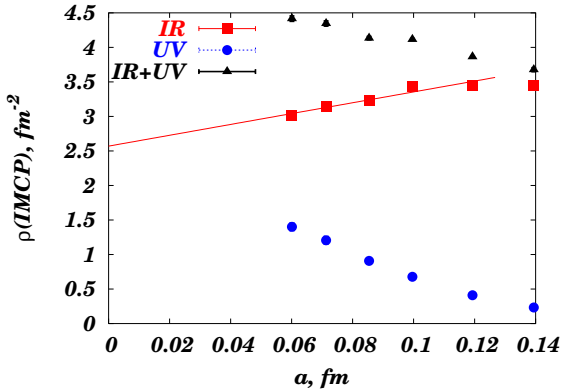}
\vspace{-5mm}
\caption{P-Vortex densities in IMCP.
    \label{fig:PVI_scaling}}
\vspace{-5mm}
\end{figure}

For P-vortices the general picture is the same, there exists one large (IR)
percolating cluster and a lot of small (UV) finite clusters. The density of
P-vortex cluster is defined as: $\rho = <P>/(N_P a^2)$, where $P$ is the
average number of plaquettes in the P-vortex cluster, $N_P= 6 L^4$ is the total
number of plaquettes on the lattice. The density of IR P-vortex cluster in
DMCP, $\rho^{IR}(DMCP)$, scales almost perfectly as is seen from
Fig.\ref{fig:PVD_scaling}. The density of IR P-vortex clusters in ICMP,
$\rho^{IR}(IMCP)$, presented on Fig.\ref{fig:PVI_scaling} depends on $a$. The
constant and linear fits $\rho^{IR}(DCMP) = \rho^{IR}_{a\to 0}(DCMP)$,
$\rho^{IR}(ICMP) = \rho^{IR}_{a\to 0}(ICMP) + Ca$ give the values of densities
which coincide within the errors in the continuum:
\begin{eqnarray}
\lefteqn{\rho^{IR}_{a\to 0}(ICMP) ~=~2.57(11)\;fm^{-2}\, ,}\\
\lefteqn{\rho^{IR}_{a\to 0}(DCMP) ~=~2.63(5)\;fm^{-2}\, ,}
\end{eqnarray}
These values can be compared with the extrapolated to the continuum density of
all clusters in ICMP ~\cite{my1}: $\rho^{IR+UV}_{a\to 0}(ICMP)\approx~4\;
fm^{-2}$.

\section{IMCP P-VORTICES AND MONOPOLES}

\begin{figure}[h]
\vspace{-5mm}
\includegraphics[scale=0.3,angle=270]{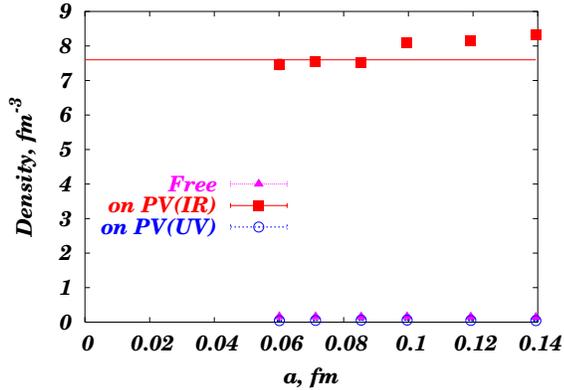}
\vspace{-5mm}
\caption{Density of IR monopole currents.
    \label{fig:MonIR_on_PV_IMC}}
\vspace{-5mm}
\end{figure}

\begin{figure}[h]
\vspace{5mm}
\includegraphics[scale=0.3,angle=270]{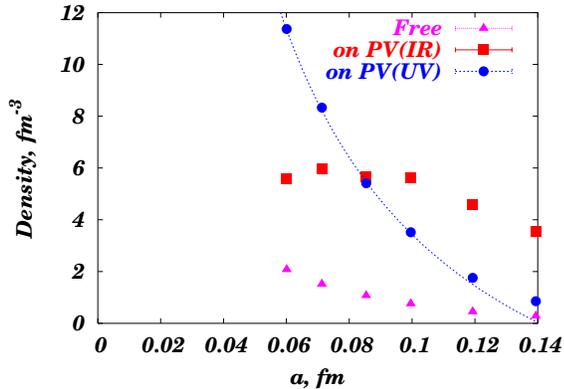}
\vspace{-5mm}
\caption{Density of UV monopole currents.
    \label{fig:MonUV_on_PV_IMC}}
\vspace{-5mm}
\end{figure}

As was observed in \cite{correlations} the main part of the monopole
trajectories lie on P-vortices.  Here we will discuss this issue in IMCP for
various values of the lattice spacing and we will discriminate between IR and
UV clusters of monopoles and P-vortices. In Figs.~\ref{fig:MonIR_on_PV_IMC},
\ref{fig:MonUV_on_PV_IMC} we present the densities of IR and UV monopole
clusters lying on IR and UV P-vortices. We also show the densities of IR and UV
monopole clusters which do not belong to P-vortices (``free''
monopoles).
We see that the density of IR monopoles lying on IR P-vortices is much
larger than other densities of IR monopoles. Similarly, the total density of UV
monopoles is mainly due to monopoles lying on UV P-vortices. This density is
divergent as $1/a$ at small values of $a$. The fit of four points at small
values of $a$ by the function $C_1+C_2/a$ gives:
\begin{equation}
C_1\approx -8.5(2)fm^{-3},\;
C_2\approx 1.19(1)fm^{-2}
\end{equation}

For DMCP we also observe the correlations of monopoles and P-vortices, and we
will publish these results elsewhere.

\section{RESULTS}
In the continuum limit IR P-vortex densities in direct and indirect maximal
center projections coincide within statistical errors. Correlations of
monopoles and vortices survive in the continuum limit, most of the IR (UV)
monopole currents belonging to IR (UV) P-vortices. 
The densities of IR P-vortices
and IR monopole currents lying on IR P-vortices are finite in the continuum,
while the densities of UV P-vortices and UV monopoles lying on UV P-vortices
diverge as $1/a$ when we approach the continuum limit.

\end{document}